\documentclass[aps,prl,twocolumn,showpacs,superscriptaddress,groupedaddress]{revtex4}  % for review and submission
\usepackage{graphicx}  % needed for figures
\usepackage[caption=false]{subfig} %for subfigures.its  caption needs to be switched off
\usepackage{floatrow}
\usepackage{dcolumn}   % needed for some tables
\usepackage{bm}        % for math
\usepackage{amssymb}   % for math
\usepackage{enumerate} %for enumerating (in roman numerals)

\begin{document}

% the following line is for submission, including submission to the arXiv!!
%\hspace{5.2in} \mbox{Fermilab-Pub-04/xxx-E}

\title{Mass-density and Phonon-frequency Relaxation Dynamics of Water and Ice at Cooling}
\author{Chang Q Sun}
 \email{ecqsun@ntu.edu.sg}
\affiliation{Key Laboratory of Low-Dimensional Materials and Application Technologies, and Faculty of Materials and Optoelectronics and Physics, Xiangtan University, Hunan 411105, China}
 \affiliation{School of Electrical and Electronic Engineering, Nanyang Technological University, Singapore 639798}

\author{Xi Zhang}
\affiliation{School of Electrical and Electronic Engineering, Nanyang Technological University, Singapore 639798}

\author{Xiaojian Fu}
\affiliation{State Key Laboratory of New Ceramics and Fine Processing, Department of Materials Science and Engineering, Tsinghua University, Beijing 100084, China}
\affiliation{ College of Materials Science and Engineering, China Jiliang University, Hangzhou 310018, China}

\author{Weitao Zheng}
\affiliation{School of Materials Science, Jilin University, Changchun 130012, China}

\author {Jer-lai Kuo}
\affiliation {Institute of Atomic and Molecular Sciences, Academia Sinica, Taipei 10617, Taiwan}

\author{Yichun Zhou}
\affiliation{Key Laboratory of Low-Dimensional Materials and Application Technologies, and Faculty of Materials and Optoelectronics and Physics, Xiangtan University, Hunan 411105, China}

\author{Zexiang Shen}
\affiliation{School of Physics, Nanyang Technological University, Singapore 639798}

\author{Ji Zhou}
\email{zhouji@Tsinghua.edu.cn}
\affiliation{State Key Laboratory of New Ceramics and Fine Processing, Department of Materials Science and Engineering, Tsinghua University, Beijing 100084, China}

\date{\today}

\begin{abstract}Coulomb repulsion between the bonding electron pair in the H-O covalent bond (denoted by``-'') and the nonbonding electron pair of O (``:") and the specific-heat disparity between the O:H and the H-O segments of the entire hydrogen bond (O:H-O) are shown to determine the O:H-O bond angle-length-stiffness relaxation dynamics and the density anomalies of water and ice. The bonding part with relatively lower specific-heat is more easily activated by cooling, which serves as the ``master" and contracts, while forcing the ``slave" with higher specific-heat to elongate (via Coulomb repulsion) by different amounts. In the liquid and solid phases, the O:H van der Waals bond serves as the master and becomes significantly shorter and stiffer while the H-O bond becomes slightly longer and softer (phonon frequency is a measure of bond stiffness), resulting in an O:H-O cooling contraction and the seemingly ``regular'' process of cooling densification. In the water-ice transition phase, the master and the slave swap roles, thus resulting in an O:H-O elongation and volume expansion during freezing. In ice, the O--O distance is longer than it is in water, resulting in a lower density, so that ice floats. 
\end{abstract}

\pacs{61.20.Ja, 61.30.Hn, 68.08.Bc\\Supplementary Information and a molecular dynamics movie are accompanied.}

\maketitle

\begin{figure}[!hbtp]
\sidesubfloat[]{\includegraphics[width=3.2in, trim=150 250 360 110, clip]{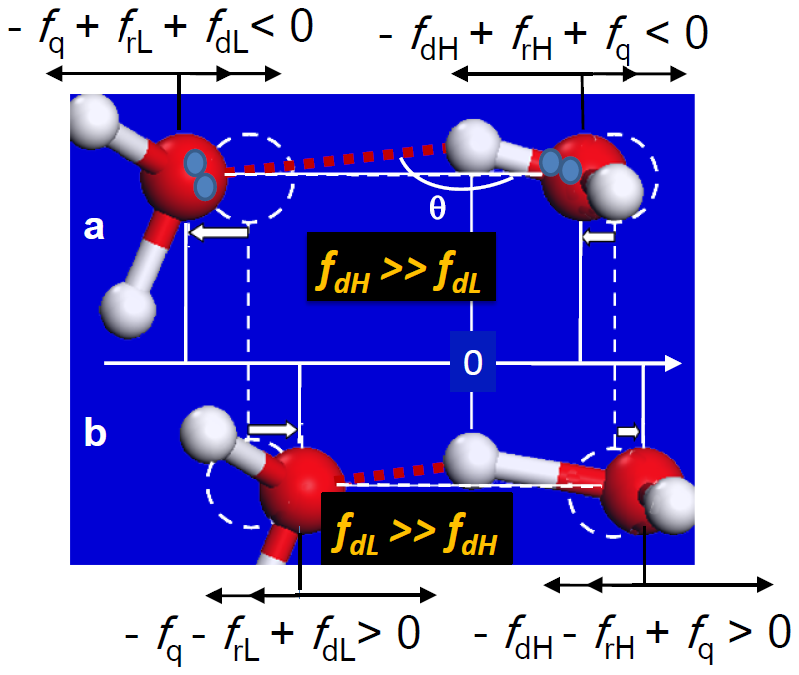}
\label{fig1a}}\\
\sidesubfloat[]{\includegraphics[width=3.0in, trim=0 0 0 30, clip]{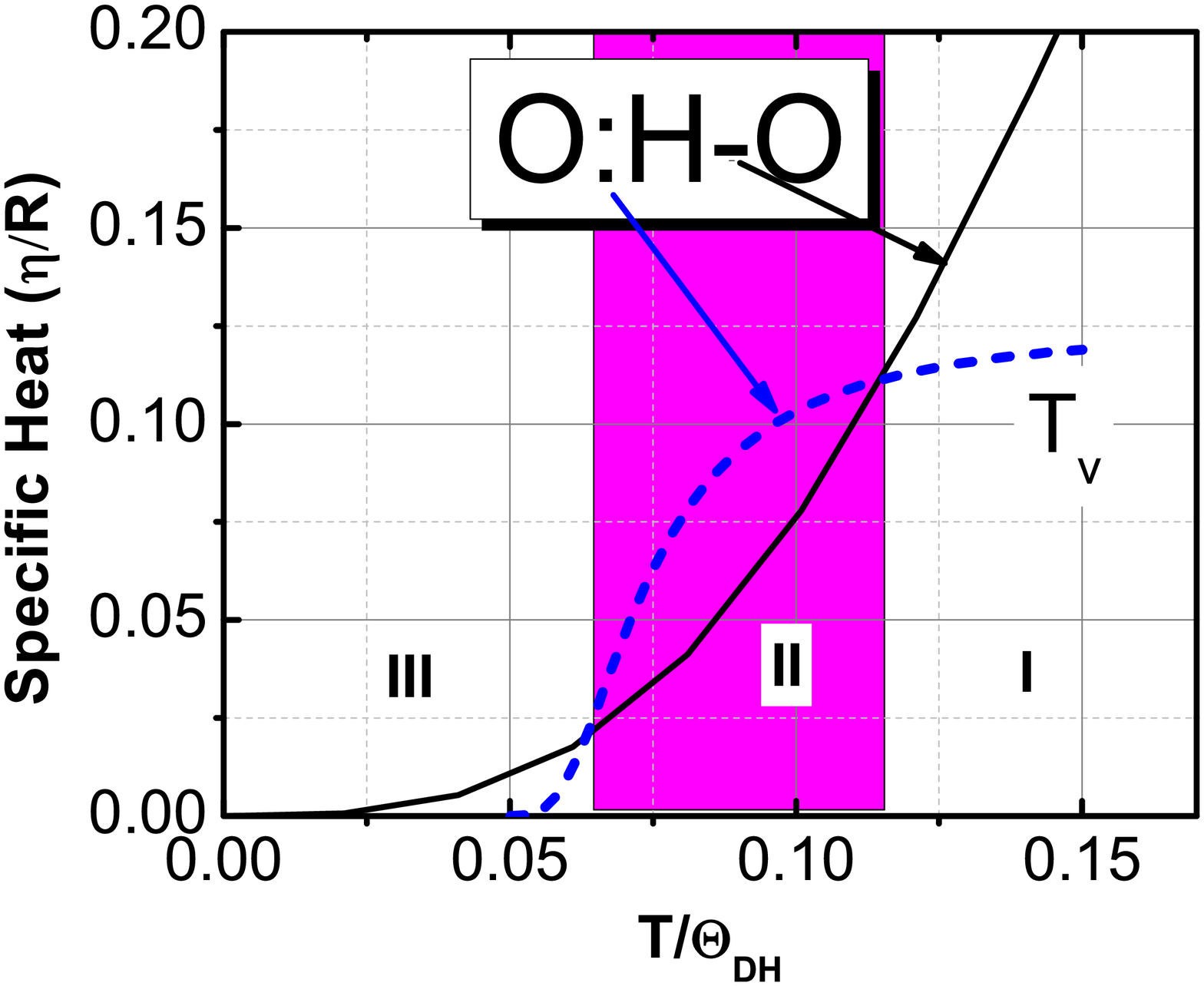}
\label{fig1b}}

\vspace{-0.8em}
\caption{(a, b) Forces and (c) the respective specific heats of the two segments in the O:H-O bond. Combined with the Coulomb repulsion $f_q$ between the electron pairs (pairs of dots in a) and the resistance to deformation $f_{rx}$ ($x = H$ for the H-O and $L$ for the O:H segment), the cooling contraction force $f_{dx}$ drives the two segments to relax in the same direction but by different amounts. (c) Because of the difference in their Debye temperatures (Table \ref{tab1}), the specific heat $\eta_L$ of O:H rises faster towards the asymptotic maximum value than the $\eta_H$. Such a specific-heat decomposition results in three regions that correspond, respectively, to the phases of liquid (I), solid (III), and liquid-solid transition (II) with different specific-heat ratios. R is the gas constant. (The $\eta_L$ in the solid phase differs from the $\eta_L$ in the liquid, which does not influence the validity of the hypothesis).}
\label{fig1}
\end{figure}

%\section{\label{sec:level1}First-level heading}
% sections are not used for PRL papers
The anomalous behavior of the density of water as it transitions to ice and its associated hydrogen bond (defined as the entire O:H-O) angle-length-stiffness cooling relaxation dynamics continue to baffle the field, despite the intensive investigations carried out in the past decades \cite{supp,clark10a,soper10,head06,clark10b,petkov12,stone07, stokely10, huang09,english11,malla07a,malla7b,mishima98,moore11,molinero,wang09}. Established database \cite{malla07a} shows that both the liquid and the solid H$_2$O undergoes the seemingly regular process of cooling densification at different rates. At the water-ice transition phase, volume expansion takes place and results in ice with a density 9$\%$ lower than the maximal density of water at 277 K \cite{malla07a,supp}. The cooling densification is associated with a redshift of the high-frequency H-O phonons $\omega_H$ ($\sim$3000 cm$^{-1}$)\cite{cross37,yoshmura} while the freezing expansion is accompanied with a blueshift of the $\omega_H$ \cite{duric11}. However, the understanding of the fundamental nature underlying the observed mass-density and phonon-frequency transition dynamics and their correlation still remains incomplete. Numerous models have been developed for explaining water's expansion upon freezing or at supercooling state. The elegant models include the thermal fluctuations in the mono-phase of tetrahedrally-coordinated structures \cite{head06,petkov12} and the mixed-phase of low- and high- density fragments with thermal modulation of the fragmental ratios \cite{huang09,wernet04}. Little attention has been paid to the mechanism for the seemingly regular process of cooling densification in both the liquid and the solid phase. The Raman shift of the low-frequency O:H ($\omega_L \sim$ 200 cm$^{-1}$) phonons in various phases has not yet been systematically characterized. Therefore, there is a great need for a consistent framework to understand the density and phonon-frequency transitions and the associated hydrogen bond angle-length-stiffness relaxation dynamics of H$_2$O through its full set of states. 
\\
\indent
In this letter, we show that we have been able to reconcile the measured mass-density \cite{malla07a,supp} and Raman-frequency transitions of water/ice based on the framework of O:H-O bond specific-heat disparity, Raman sectroscopy measurements, and molecular dynamics (MD) calculations of the hydrogen bond angle-length-stiffness relaxationin of water/ice over the full temperature range. The often overlooked Coulomb repulsion between the electron pair in the H-O covalent bond and the nonbonding electron lone pair of oxygen \cite{sun12} and the specific-heat disparity of the O:H-O bond are shown to be the key to resolving the density puzzles.
\\
\indent
We consider the basic structural unit of O$^{\delta -}$:H$^{\delta +}$-O$^{\delta -}$ \cite{sun12,sunprl} (also denoted as ``O$\cdots$H-O") to represent the O$^{\delta -}$-O$^{\delta -}$ interactions in H$_2$O, except for H$_2$O under extremely high pressures and temperatures \cite{wang11}. The fraction $\delta$ represents the polarity of the H$^{\delta +}$-O$^{\delta -}$ polar-covalent bond. In Fig.\ref{fig1a}, the pair of dots on the left represents the nonbonding lone pair ``:" and the pair on the O atom on the right represents the bonding electron pair ``-". The H atom serves as the point of reference in the O:H-O system. The lone pair on the left belongs to the sp$^3$-orbit hybridized oxygen and the bonding pair is shared by the H-O and centred at sites close to oxygen. For completeness, we define the entire hydrogen bond to be O:H-O, the intra-molecular polar-covalent bond as the H-O bond and the inter-molecular van der Waals (vdW) bond as the O:H bond hereon. 

%%insert figure 1 (a,b,c)
\floatsetup[figure]{style=plain,subcapbesideposition=top}

As illustrated in Fig.\ref{fig1a}, a hydrogen bond is comprised of the O:H bond (broken lines) and the H-O bond rather than either of them alone. The H-O bond is much shorter, stronger, and stiffer than the O:H bond (comparison shown in Table \ref{tab1}). The O:H bond breaks at the evaporating point (T$_v$) of water (373 K)\cite{cross37}. However, the H-O bond is much harder to break as the bond energy of $\approx$4.0 eV \cite{supp} is twice that of the C-C bond in diamond (1.84 eV) \cite{kittel}. 

%insert table
\begin{table*}[!htbp]
\caption{Summary of the segmental length d$_x$, strength E$_x$ (energy)\cite{supp}, Debye temperature $\Theta_{Dx}$ \cite{zhao07,suter06}, stiffness $\omega_{x}$ (vibration frequency), melting point T$_{mx}$, and the inter-atomic and inter-electron-pair interactions of the O:H-O bond compared with those of the C-C bond in a diamond \cite{kittel}. 
}
\centering
\begin{tabular}{ccccccc}
\hline\hline
Segment (x) & d$_x$ (nm) & E$_x$ (eV) & $\omega_x$ (cm$^{-1}$) & $\Theta_{Dx}$ (K) & T$_{mx}$ (K) & Interaction \\
\hline
H-O & $\sim$0.10 & $\sim$4.00 & $>$3000 & $>$3000 & - & Exchange \\
O:H & $\sim$0.20 & $\sim$0.05-0.10 & $\sim$200 & 198 & 373 & vdW \\
O--O & - & - & - & - & - & Repulsion \\
C-C & 0.15 & 1.84 & 1331 & 2230 & 3800 & Exchange \\
\hline
\end{tabular}
\label{tab1}
\end{table*}

Combined with the forces of the Coulomb repulsion, $f_q$$\propto$(d$_{O-O}$)$^{-2}$, and resistance to deformation, $f_{rx}$($x = H$ for the H-O and $L$ for the O:H bond), the force of cooling contraction, $f_{dx}$ drives these two segments to relax in the same direction but by different amounts. The $k_x$, which varies nonlinearly with the $d_x$, approximates the second derivative of the inter-atomic potential at equilibrium. The magnitude of the $f_{dx}$ varies with the specific heat of the specific bond in a particular temperature range. The Coulomb repulsion between the electron pairs, as represented by the pairs of dots in Fig.\ref{fig1a}, is the key to the O:H-O bond relaxation under excitation \cite{supp,sun12}.
\\
\indent
Generally, the specific heat is regarded as a macroscopic quantity integrated over all bonds, and is the amount of energy required to raise the temperature of the substance by 1 K. However, in dealing with the representative bond of the entire specimen, one has to consider the specific heat per bond that is obtained by dividing the bulk specific heat by the total number of bonds involved. In the case of the O:H-O bond, we need to consider the specific heat ($\eta_x$) characteristics of the two segments separately (see Fig.\ref{fig1b}) because of the difference in their strengths. The slope of the specific-heat curve is determined by the Debye temperature ($\Theta_{Dx}$) while the integration of the specific-heat curve from 0 K to the T$_{mx}$ \cite{sun09} is determined by the cohesive energy of the bond energy E$_x$. The specific-heat curve of the segment with a relatively lower $\Theta_{Dx}$ value will rise to the maximum value faster than the other segment. The $\Theta_{Dx}$, which is lower than the T$_{mx}$, is proportional to the characteristic frequency of the vibration ($\omega_x$) of the segment. Thus, we have the following relations (see Table \ref{tab1}):
\begin{equation}
\label{eq1}
\left\{
\begin{array}{l}
\frac{\Theta_{DL}}{\Theta_{DH}} \approx \frac{198}{\Theta_{DH}} \approx \frac{\omega_L}{\omega_H} \approx \frac{200}{3000} \approx \frac{1}{15} \\
\frac{\left(\int^{T_{mH}}_0 \eta_H dt \right)}{\left(\int^{T_{mL}}_0 \eta_L dt \right)} \approx \frac{E_H}{E_L} \approx \frac{4.0}{0.1} \approx 40
\end{array}
\right.
\end{equation}

Based on this consideration, the maximal specific-heat ratio is estimated to be $\frac{\eta_H}{\eta_L}$$\approx$8. Such a specific-heat disparity between the O:H and the H-O segments creates three temperature regions with different $\frac{\eta_L}{\eta_H}$ ratios, which should correspond to the phases of liquid (I), solid (III), and liquid-solid transition (II). 
\\
\indent
The consistency in the number of regions (i.e. phases I, II, III) of the proposed specific-heat curve (Fig.\ref{fig1b}), the mass-density transition \cite{malla07a}, and the O:H-O bond angle-length-stiffness relaxation dynamics (Fig. 2 and Fig. 3) suggest that the segment with a relatively lower specific heat is thermally more active than the other segment. This thermally active segment serves as the ``master'' that undergoes cooling contraction while forcing the other ``slave'' segment to elongate through Coulomb repulsion. Therefore, as can be derived from Fig.\ref{fig1b}, the specific-heat ratio, the master segment, and the O--O length change in each temperature region are correlated as follows (see \cite{supp} for details):

\vspace{-1.0em}
\begin{widetext}
\begin{equation}
\label{eq2}
\left.
\begin{array}{l}
\text{II} \\
\text{I, III} \\
\text{Transition}
\end{array}
\begin{array}{l}
(\eta_H < \eta_L): \\
(\eta_L < \eta_H): \\
(\eta_H = \eta_L):
\end{array}
\begin{array}{l}
f_{dH}>(f_{dL} + f_{rL} + f_{rH}) \\
f_{dL}>(f_{dH} + f_{rL} + f_{rH})\\
f_{dH}=f_{dL} 
\end{array}
\right\}
\Rightarrow \Delta d_{O-O} \Rightarrow \left(\Delta V \right)^{1/3}
\left\{
\begin{array}{l}
<\\
>\\
=
\end{array}
\right\}
0
\end{equation}
\end{widetext}
\noindent
Because of the strength difference of the two segments \cite{sun12}, the length of the softer O:H bond always relaxes more than that of the stiffer H-O bond: $|\Delta d_L|$$>$$|\Delta d_H|$. The two segments relax in the same direction because of the repulsion. Thus, we expect the O:H-O bond to relax in the following manners during cooling:\\
i) In the transition phase II, $\eta_H$$<$$\eta_L$ and $f_{dH}$$\gg$$f_{dL}$. The H-O bond contraction dominates. The stiffer H-O bond contracts less than the O:H bond elongates, resulting in $\Delta d_{O-O}$=$\Delta d_L - \Delta d_H$$>$0. Therefore, a net O--O length gain and an accompanying volume expansion ($\Delta$V $>$ 0) takes place. \\
ii) In the liquid I and the solid III phase, $\eta_L$$<$$\eta_H$ and $f_{dL}$$\gg$$f_{dH}$. The master and the slave swap roles. The softer O:H bond contracts significantly more than the H-O bond elongates, $\Delta d_{O-O}$=$\Delta d_H - \Delta d_L$$<$0. Hence, a net O--O contraction results in a gain in the mass density. \\
iii) At the crossing points (Fig.\ref{fig1b}), $\eta_H$$=$$\eta_L$ and $f_{dH}$=$f_{dL}$. There is a transition between O--O expansion and contraction, corresponding to the density maximum at 277 K and the density minimum below the freezing point \cite{malla07a,malla7b}. \\
iv) Meanwhile, the repulsion increases the O:H-O angle $\theta$ and polarizes the electron pairs during relaxation. \\
\indent
It has been shown that a segment increases in stiffness as it becomes shorter, while the opposite occurs as it elongates \cite{sun12,li12}. The Raman shift, which is proportional to the square root of bond stiffness, approximates the length and strength change of the bond during relaxation directly. Approximating the vibration energy of a vibration system to the Taylor series of the inter-atomic potential energy, u$_x$(r), leads to:
\vspace{-0.3em}
\begin{equation}
\label{eq3}
\Delta \omega_x \propto \left( \frac{\partial^2 u_x (r)}{\mu \partial r^2} \bigg|_{r=d_x} \right)^{1/2} \propto \frac{\sqrt{E_x/ \mu}}{d_x} \propto \sqrt{Y_x d_x}
\end{equation}

\noindent
The stiffness is the product of the Young's modulus (Y$_x$$\propto$E$_x$/d$^3_x$) and the length of the segment in question \cite{sun12}. The $\mu$ is the reduced mass of the vibrating dimer. Therefore, the Raman shift is a measure of the segmental stiffness.
\\
\indent
In order to verify our hypotheses and predictions regarding the O:H-O bond angle-length-stiffness change, the specific-heat disparity, and the density and phonon-frequency anomalies of water/ice, we have conducted Raman measurements and MD calculations as a function of temperature. Two MD computational methods were used in examining the mono- and the mixed-phase models. Details of the experimental and calculation procedures are described in the supplementary information \cite{supp}. 

%insert figure 2(a,b,c,d)
\floatsetup[figure]{style=plain,subcapbesideposition=top}
\begin{figure}[!hbtp]
\sidesubfloat[]{\includegraphics[width=1.55in, trim=0 50 40 20, clip]{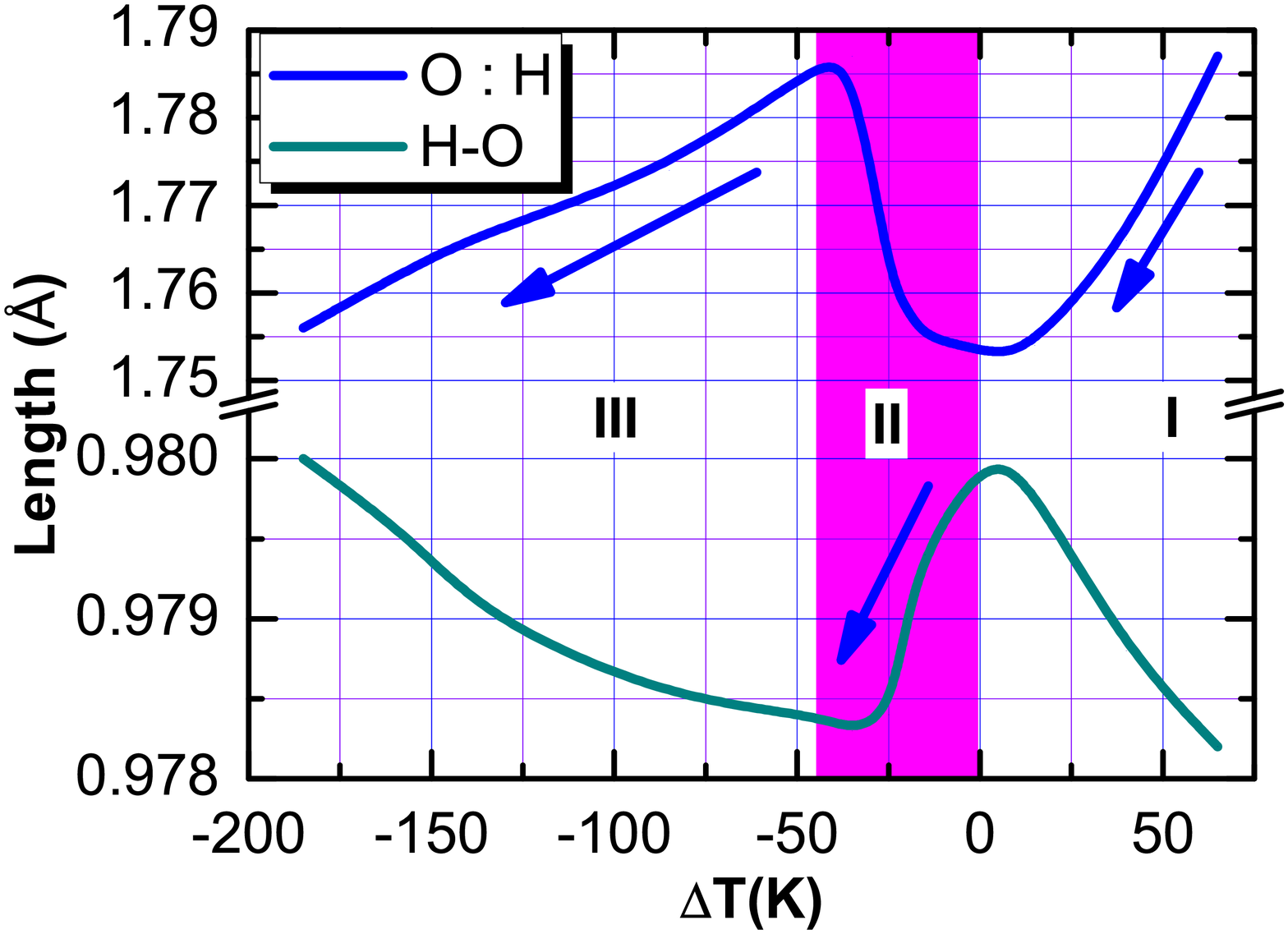}
\label{fig2a}}
\sidesubfloat[]{\includegraphics[width=1.6in, trim=10 18 00 20, clip]{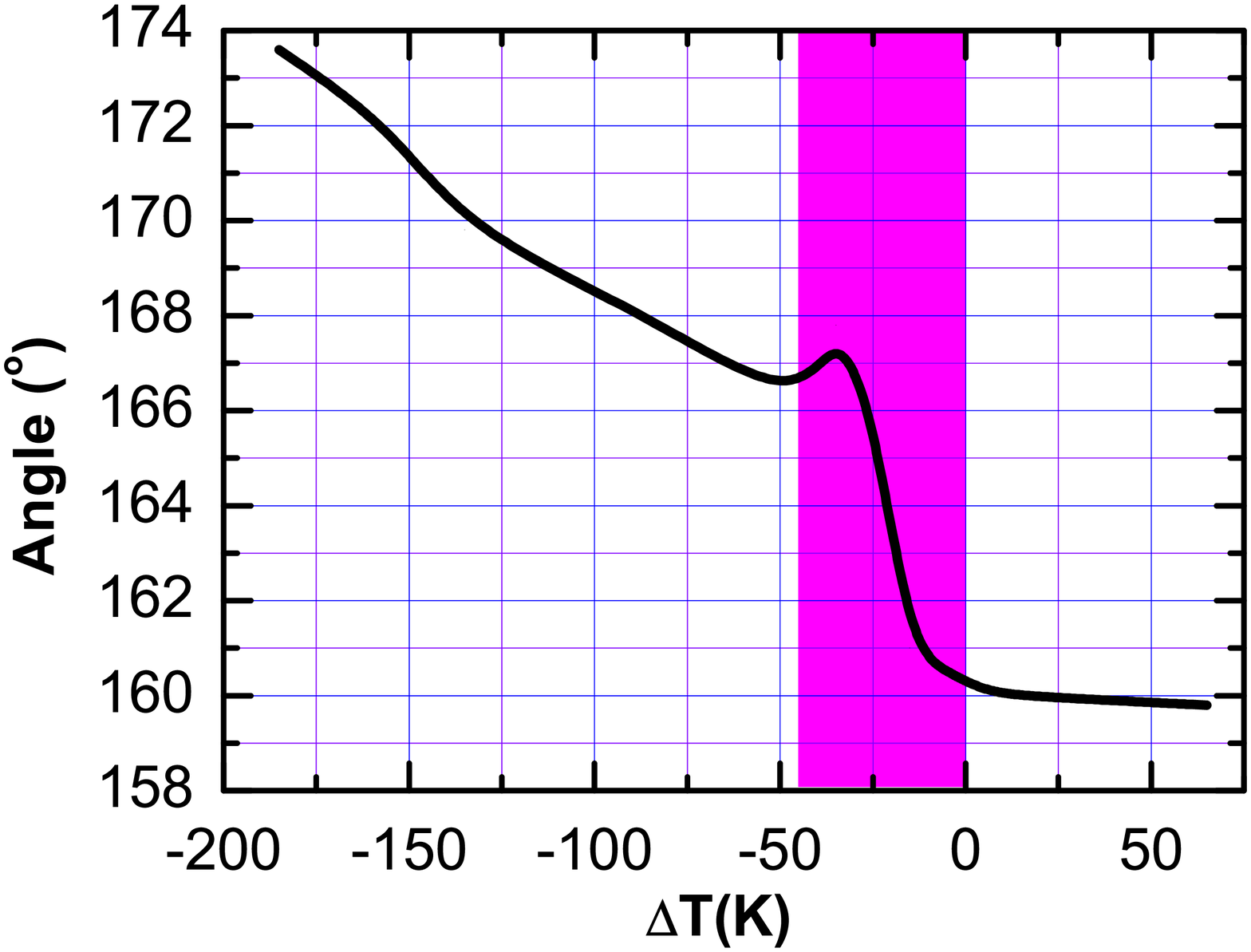}
\label{fig2b}}\\
\sidesubfloat[]{\includegraphics[width=1.45in, trim=220 600 200 100, clip]{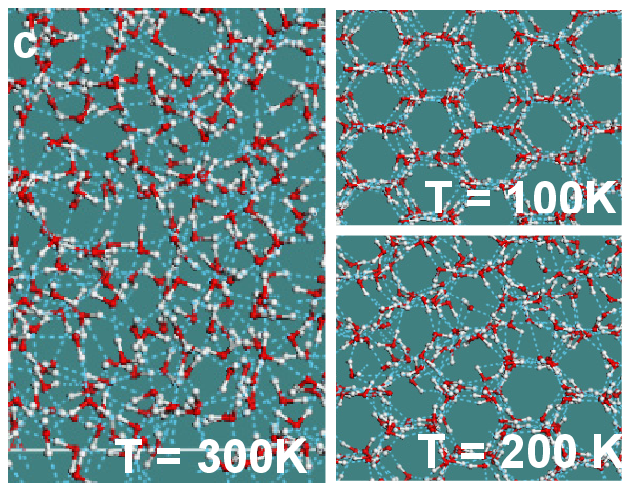}
\label{fig2c}}
\sidesubfloat[]{\includegraphics[width=1.5in, trim=0 10 115 50, clip]{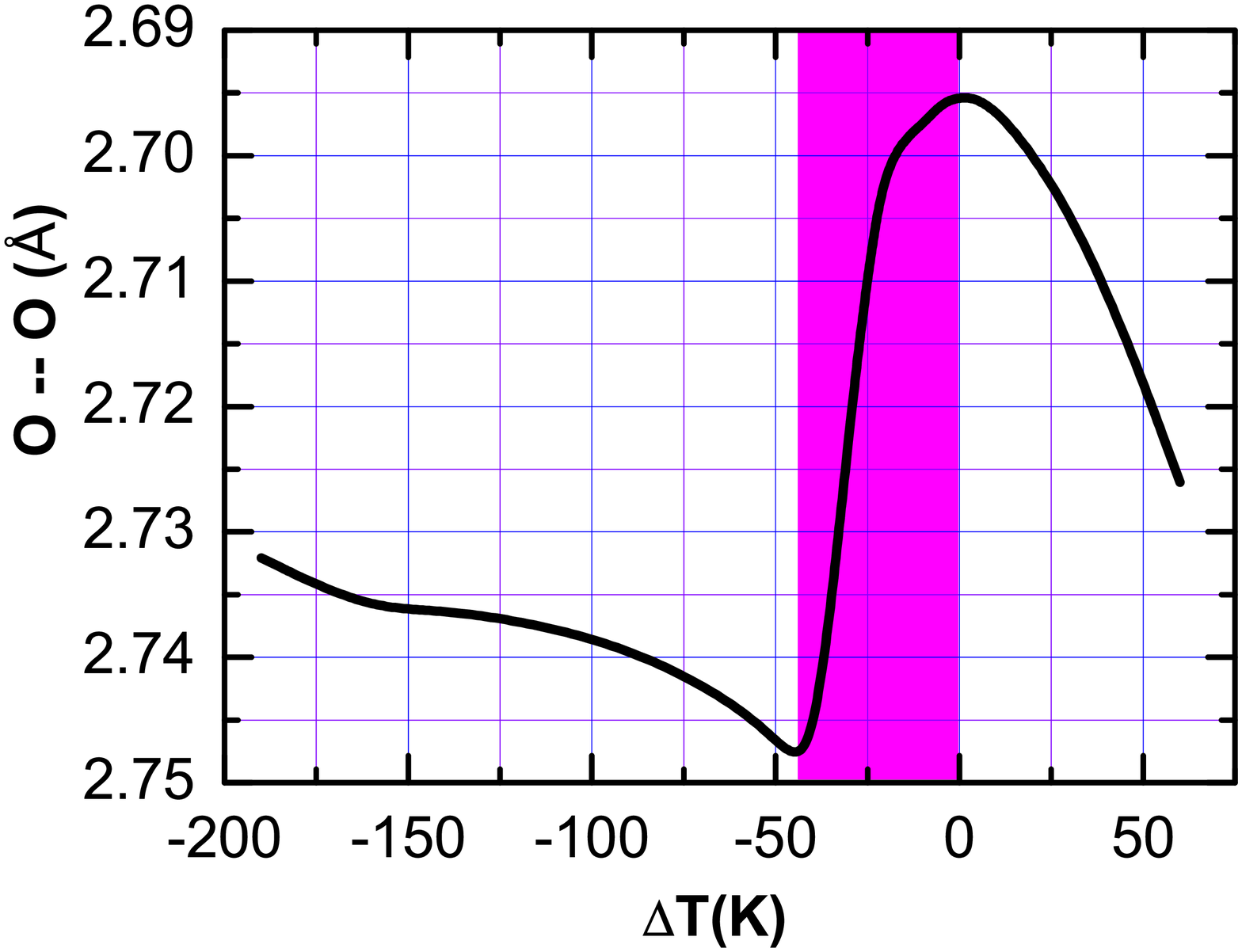}
\label{fig2d}}\\
\caption{(a) Segmental length change of the O:H-O bond in the phases of liquid (I), solid (III), and liquid-solid transition (II). Arrows denote the cooling contraction of the master segments, which are coupled with the expansion of the slaves. $\Delta T$ = T - T$_{max}$ with T$_{max}$ = 277 K is the maximal density temperature. (b) O:H-O bond angle widening driven by cooling also exhibits three regions. (c) Snapshots of the MD trajectory show that the V-shaped H-O-H molecues remain intact at 300 K because of the robustness of the H-O bond ($\sim$4.0 eV/bond) with pronounced quantum fluctuations in the angle and in the d$_L$ in liquid phase. (d) The change of the O--O distance agrees with the measured three-region water and ice densities \cite{supp,malla07a}. In ice, the O--O distance is longer than that in water, which results in ice floating.}
\label{fig2}
\end{figure}

Fig.\ref{fig2} shows the MD-derived change of (a) the H-O bond and the O:H bond length, (b) the O:H-O bond angle $\theta$, (c) the snapshots of the MD trajectory, and (d) the O--O distance as a function of temperature. As shown in Fig.\ref{fig2a}, the shortening of the master segments (denoted with arrows) is always coupled with a lengthening of the slaves during cooling. The temperature range of interest consists of three regions: in the liquid region I and the solid region III, the O:H bond contracts significantly more than the H-O bond elongates, resulting in a net loss of the O--O length. Thus, cooling-driven densification of H$_2$O happens in both the liquid and the solid phase. This mechanism differs completely from the mechanism conventionally adopted for the standard cooling densification of other regular materials in which only one kind of chemical bond is involved. In other materials, cooling shortens and stiffens all the inter-atomic bonds \cite{sun05}. In contrast, in the transition phase II \cite{malla07a,mishima98,moore11}, the master and the slave swap roles. The O:H bond elongates more than the H-O bond shortens so that a net gain in the O--O length occurs. 
\\
\indent
The $\theta$ angle widening (Fig.\ref{fig2b}) could also contribute to the volume change. In the liquid phase I, the mean $\theta$ valued at 160$^\circ$ remains almost constant. However, the snapshots of the MD trajectory in Fig.\ref{fig2c} and the MD movie in the attached \cite{MD movie} show that the V-shaped H-O-H molecules remain intact at 300 K over the entire duration recorded, accompanied by high fluctuations in the $\theta$ and the $d_L$ in this regime, which indicates the dominance of tetrahedrally-coordinated water molecules \cite{kuhne13}. In region II, cooling widens the $\theta$ from 160$^\circ$ to 167$^\circ$, which contributes a maximum of +1.75$\%$ to the O:H-O bond elongation and ~5.25$\%$ to volume expansion. In phase III, the $\theta$ increases from 167$^\circ$ to 174$^\circ$ and this trend results in a maximal value of -2.76$\%$ to the volume contraction. 
\\
\indent
The calculated temperature dependence of the O--O distance as shown in Fig.\ref{fig2d} matches satisfactorily with that of the measured density profile \cite{supp,malla07a}. Importantly, the O--O distance is longer in ice than it is in water, and therefore, ice floats. 

%%insert figure 3 (a,b) 
\floatsetup[figure]{style=plain,subcapbesideposition=top}
\begin{figure}[!hbtp]
\sidesubfloat[]{\includegraphics[width=1.5in, trim=95 25 62 90, clip]{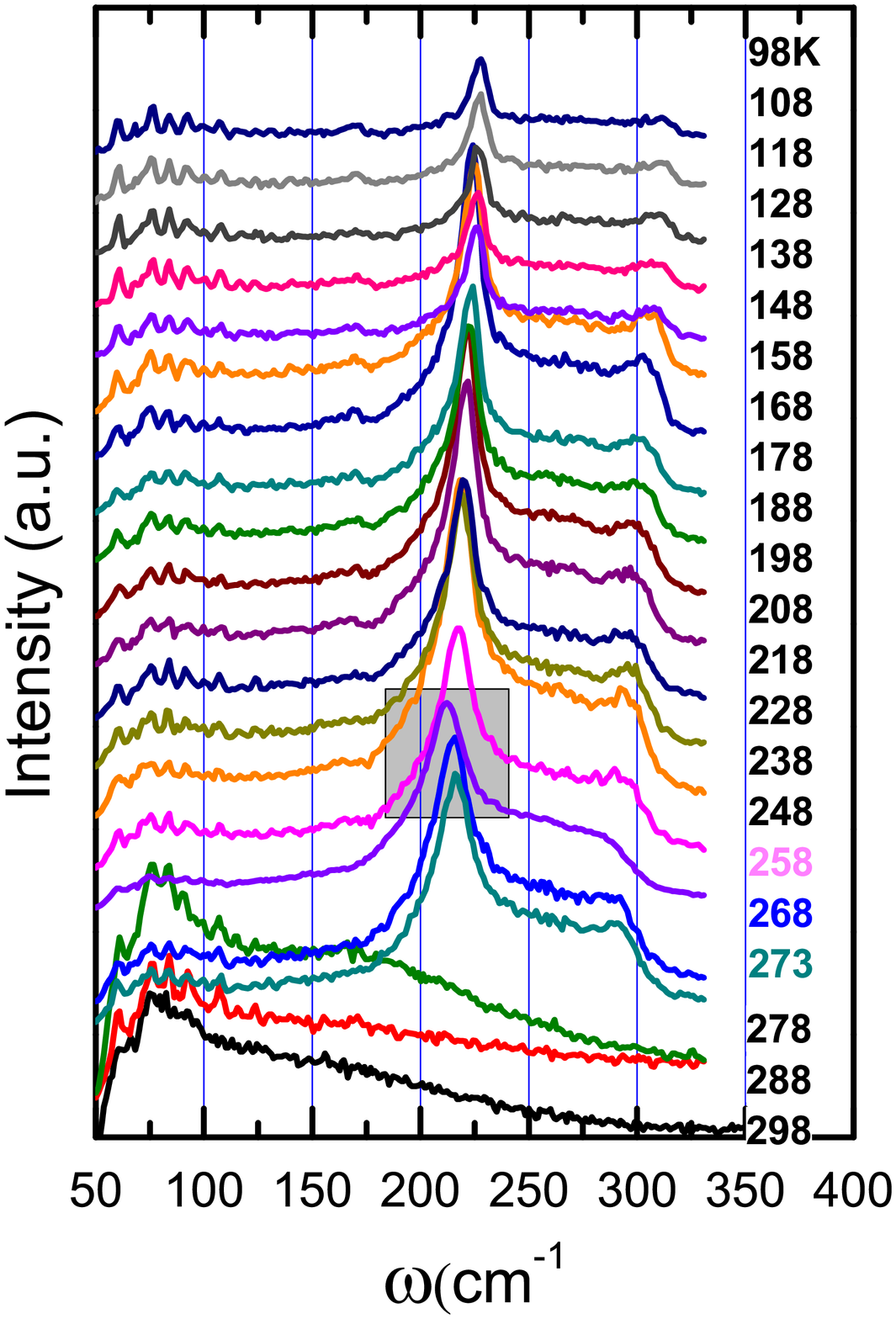}
\label{fig3a}}
\sidesubfloat[]{\includegraphics[width=1.46in, trim=105 25 60 100, clip]{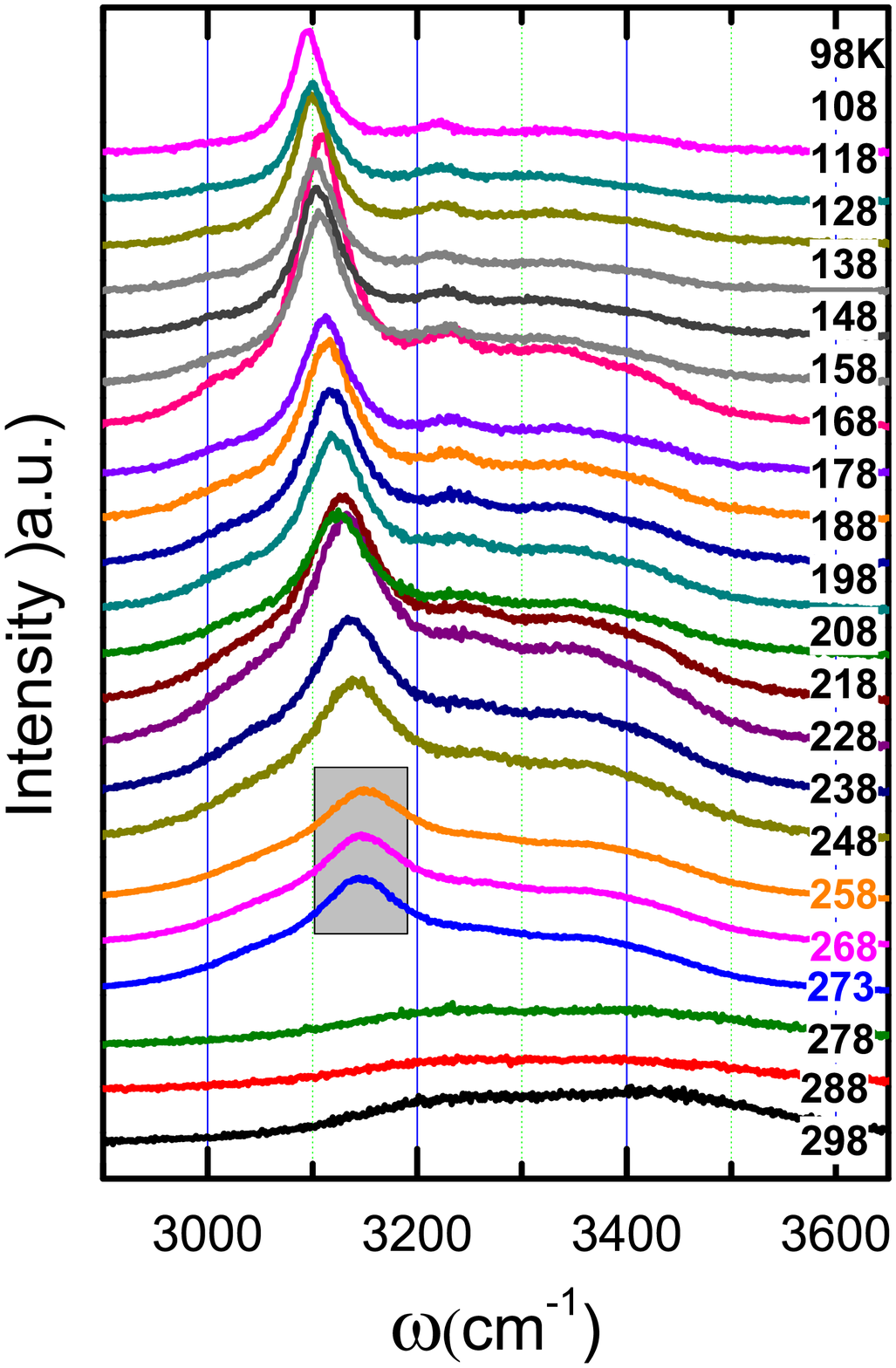}
\label{fig3b}}
\vspace{-0.8em}
\caption{Temperature dependent Raman shifts of (a) $\omega_L$ (the O:H phonon) and (b) $\omega_H$ show three regions: T$>$ 273 K (I), 273 K$<$T$<$258 K, and T$<$258 K (III) .}
\label{fig3}
\end{figure}

\indent
The measured Raman spectra in Fig.\ref{fig3} show three regions: T$>$273 K (I), 273 $\geq$ T $\geq$ 258 K (II), and T$<$258 K (III), which are in agreement with the MD calculations \cite{supp} and our predictions: \\
i) At T$>$273 K (I), abrupt shifts of the $\omega_L$ from 75 to 220 cm$^{-1}$ and the $\omega_H$  from 3200 to 3150 cm$^{-1}$ indicate ice formation. The coupled $\omega_L$ blueshift and $\omega_H$  redshift indicate that cooling shortens and stiffens the O:H bond but lengthens and softens the H-O bond in the liquid phase, which confirms the predicted master role of the O:H bond. \\
ii) At T$<$258 K (III), the trend of phonon relaxation remains the same as it is in the region of T $>$ 273 K despite a change in the relaxation rates. Cooling from 258 K stiffens the $\omega_L$ from 215 to 230 cm$^{-1}$ and softens the $\omega_H$ from 3170 to 3100 cm$^{-1}$. Other supplementary peaks at $\sim$300 and $\sim$3400 cm$^{-1}$ are found to be insignificant. The cooling softening of the $\omega_H$ mode agrees with that measured using IR spectroscopy \cite{medcrafy13} of ice clusters of 8$\sim$150 nm sizes. When the temperature drops from 209 to 30 K the $\omega_H$ shift from 3253 to 3218 cm$^{-1}$. For clusters of 5 nm size or smaller, the $\omega_H$ shifts by an addition of 40 cm$^{-1}$. \\
iii) At 273$\sim$258 K (II), the situation reverses. Cooling shifts the $\omega_H$ from 3150 to 3170 cm$^{-1}$ and the  $\omega_L$ from 220 to 215 cm$^{-1}$, see the shaded areas. Agreeing with the Raman $\omega_H$ shift measured in the temperature range between 270 and 273 K \cite{supp,duric11}, the coupling of the $\omega_H$ blueshift and the $\omega_L$ redshift confirms the exchange in the master and the slave role of the O:H and the H-O bond during freezing. 
\\
\indent
The MD-movie \cite{MD movie} shows that in the liquid phase, the H$^{\delta +}$ and the O$^{\delta -}$ attract each other between the H$^{\delta +}$:O$^{\delta-}$  but the O$^{\delta -}$-O$^{\delta -}$ repulsion prevents this occurrence. The intact O-H-O motifs are moving restlessly because of the high fluctuations and frequent switching of the H$^{\delta  +}$:O$^{\delta -}$ interactions. Furthermore, the coupled cooling $\omega_L$ blueshift and $\omega_H$ redshift provide further evidence for the persistence of the Coulomb repulsion between the bonding and the nonbonding electron pairs in liquid. The presence of the electron lone pair results from the sp$^3$-orbit hybridization of oxygen that tends to form tetrahedral bonds with its neighbors \cite{supp,kuhne13}. Therefore, the H$¬_2$O in the bulk form of liquid could possesses the tetrahedrally-coordinated structures with thermal fluctuation \cite{petkov12,kuhne13,pauling35}. Snapshots of the MD trajectory in \cite{supp} revealed little discrepancy between the mono- and the mixed-phase structural models. 
\\
\indent
The proposed mechanisms for: i) the seemingly regular processes of cooling densification of the liquid and the solid H$_2$O, ii) the abnormal freezing expansion, iii) the floating of ice, and, iv) the three-region O:H-O bond angle-length-stiffness relaxation dynamics of water and ice have been justified. Agreement between our calculations and the measured mass-density \cite{malla07a} and phonon-frequency relaxation dynamics in the temperature range of interest has verified our hypotheses and predictions:\\
i) Inter-electron-pair Coulomb repulsion and the segmental specific-heat disparity of the O:H-O bond determine the change in its angle, length and stiffness and the density and the phonon-frequency anomalies of water ice. \\
ii) The segment with a relatively lower specific-heat contracts and drives the O:H-O bond cooling relaxation. The softer O:H bond always relaxes more in length than the stiffer H-O bond does in the same direction. The cooling widening of the O:H-O angle contributes positively to the volume expansion at freezing.\\
iii) In the liquid and the solid phase, the O:H bond contracts more than the H-O bond elongates, resulting in the cooling densification of water and ice, which is completely different from the process experienced by other regular materials. \\
iv) In the freezing transition phase, the H-O bond contracts less than the O:H bond lengthens, resulting in expansion during freezing. \\
v) The O--O distance is larger in ice than it is in water, and therefore, ice floats. \\
vi) The segment increases in stiffness as it shortens, while the opposite occurs as it elongates. The density variation of water ice is correlated to the incoporative O:H and H-O phonon-frequency relaxaion dynamics.

Special thanks to Phillip Ball, Yi Sun, Buddhudu Srinivasa, and John Colligon for their comments and expertise. Financial support received from NSF (Nos.: 21273191, 1033003, and 90922025) China is gratefully acknowledged.


\begin{thebibliography}{99}

\bibitem{supp}
Supplementary Information.
\bibitem{clark10a}
G. N. I. Clark, C. D. Cappa, J. D. Smith, R. J. Saykally, and T. Head-Gordon, Mol. Phys. 108, 1415 (2010).
\bibitem{soper10}
A. K. Soper, J. Teixeira, and T. Head-Gordon, PNAS 107, E44 (2010).
\bibitem{head06}
T. Head-Gordon, and M. E. Johnson, PNAS 103, 7973 (2006).
\bibitem{clark10b}
G. N. Clark, G. L. Hura, J. Teixeira, A. K. Soper, and T. Head-Gordon, PNAS 107, 14003 (2010).
\bibitem{petkov12}
V. Petkov, Y. Ren, and M. Suchomel, J Phys: Condens Matter 24, 155102 (2012).
\bibitem{stone07}
A. J. Stone, Science 315, 1228 (2007).
\bibitem{stokely10}
K. Stokely, M. G. Mazza, H. E. Stanley, and G. Franzese, PNAS 107, 1301 (2010).
\bibitem{huang09}
C. Huang, K. T. Wikfeldt, T. Tokushima, D. Nordlund, Y. Harada, U. Bergmann, M. Niebuhr, T. M. Weiss, Y. Horikawa, M. Leetmaa, M. P. Ljungberg, O. Takahashi, A. Lenz, L. Ojamäe, A. P. Lyubartsev, S. Shin, L. G. M. Pettersson, and A. Nilsson, PNAS 106, 15214 (2009).
\bibitem{english11}
N. J. English, and J. S. Tse, Phys. Rev. Lett. 106, 037801 (2011).
\bibitem{malla07a}
F. Mallamace, C. Branca, M. Broccio, C. Corsaro, C. Y. Mou, and S. H. Chen, PNAS 104, 18387 (2007).
\bibitem{malla7b}
F. Mallamace, M. Broccio, C. Corsaro, A. Faraone, D. Majolino, V. Venuti, L. Liu, C. Y. Mou, and S. H. Chen, PNAS 104, 424 (2007).
\bibitem{mishima98}
O. Mishima, and H. E. Stanley, Nature 396, 329 (1998).
\bibitem{moore11}
E. B. Moore, and V. Molinero, Nature 479, 506 (2011).
\bibitem{molinero}
V. Molinero, and E. B. Moore, J. Phys. Chem. B 113, 4008 (2009).
\bibitem{wang09}
C. Wang, H. Lu, Z. Wang, P. Xiu, B. Zhou, G. Zuo, R. Wan, J. Hu, and H. Fang, Phys. Rev. Lett. 103, 137801 (2009).
\bibitem{cross37}
P. C. Cross, J. Burnham, and P. A. Leighton, J. Am. Chem. Soc. 59, 1134 (1937).
\bibitem{yoshmura}
Y. Yoshimura, S. T. Stewart, H. K. Mao, and R. J. Hemley, J. Chem. Phys. 126, 174505 (2007).
\bibitem{duric11}
I. Durickovic, R. Claverie, P. Bourson, M. Marchetti, J. M. Chassot, and M. D. Fontana, J. Raman Spec. 42, 1408 (2011).
\bibitem{wernet04}
P. Wernet, D. Nordlund, U. Bergmann, M. Cavalleri, M. Odelius, H. Ogasawara, L. A. Naslund, T. K. Hirsch, L. Ojamae, P. Glatzel, L. G. M. Pettersson, and A. Nilsson, Science 304, 995 (2004).
\bibitem{sun12}
C. Q. Sun, X. Zhang, and W. T. Zheng, Chem Sci 3, 1455 (2012).
\bibitem{sunprl}
C. Q. Sun, X. Zhang, J. Zhou, Y. L. Huang, Y. Zhou, C., and W. T. Zheng, Phys. Rev. Lett. (LW13133, revised).
\bibitem{wang11}
Y. Wang, H. Liu, J. Lv, L. Zhu, H. Wang, and Y. Ma, Nat Commun 2, 563 (2011).
\bibitem{kittel}
C. Kittel, Intrduction to Solid State Physics (John Willey and Sons, New York, 2005).
\bibitem{zhao07}
M. Zhao, W. T. Zheng, J. C. Li, Z. Wen, M. X. Gu, and C. Q. Sun, Phys Rev B 75, 085427 (2007).
\bibitem{suter06}
M. T. Suter, P. U. Andersson, and J. B. Pettersson, J Chem Phys 125, 174704 (2006).
\bibitem{sun09}
C. Q. Sun, Prog. Mater Sci. 54, 179 (2009).
\bibitem{sun05}
C. Q. Sun, L. K. Pan, C. M. Li, and S. Li, Phys Rev B 72, 134301 (2005).
\bibitem{li12}
J. W. Li, S. Z. Ma, X. J. Liu, Z. F. Zhou, and C. Q. Sun, Chem. Rev. 112, 2833 (2012).
\bibitem{MD movie}
MD movie showing the quantum fluctuating of the tetrhdral molecules at 300 K.
\bibitem{kuhne13}
T. D. Kühne, and R. Z. Khaliullin, Nat Commun 4, 1450 (2013).
\bibitem{pauling35}
L. Pauling, Journal of the American Chemical Society 57, 2680 (1935).
\bibitem{medcrafy13}
C. Medcraft, D. McNaughton, C. D. Thompson, D. R. Appadoo, S. Bauerecker, and E. G. Robertson, PCCP 15, 3630 (2013).


\end{thebibliography}
\end{document}